\documentclass[journal = jpccck, manuscript = letter]{achemso}
\usepackage{achemso}
\setkeys{acs}{usetitle = true}
\usepackage{graphicx,epsf,epsfig}
\usepackage{dcolumn}
\usepackage{bm}
\usepackage{color}
\usepackage{subfig}
\usepackage{upgreek}
\usepackage{epstopdf}
\usepackage{url}
\usepackage{amsmath}
\usepackage{amssymb}
\usepackage{breakurl}
\usepackage{braket}
\usepackage{lmodern}
\usepackage[T1]{fontenc}
\setkeys{acs}{
  usetitle = true,
  etalmode = truncate,
  maxauthors = 10
}

\title[\texttt{achemso}]
{Improved Energy Pooling Efficiency Through Inhibited Spontaneous Emission}  

\author{Michael D. LaCount}
\affiliation{Department of Physics, Colorado School of Mines, Golden, CO 80401, USA}
\author{Mark T. Lusk}
\email{mlusk@mines.edu}
\affiliation{Department of Physics, Colorado School of Mines, Golden, CO 80401, USA}

\begin{document}

\begin{abstract}
The radiative lifetime of molecules or atoms can be increased by placing them within a tuned conductive cavity that inhibits spontaneous emission. This was examined as a possible means of enhancing three-body, singlet-based upconversion, known as energy pooling. Achieving efficient upconversion of light has potential applications in the fields of photovoltaics, biofuels, and medicine. The affect of the photonically constrained environment on pooling efficiency was quantified using a kinetic model populated with data from molecular quantum electrodynamics, perturbation theory, and ab initio calculations. This model was applied to a system with fluorescein donors and a hexabenzocoronene acceptor. Placing the molecules within a conducting cavity was found to increase the efficiency of energy pooling by increasing both the donor lifetime and the acceptor emission rate---i.e. a combination of inhibited spontaneous emission and the Purcell effect. A model system with a free-space pooling efficiency of 23\% was found to have an efficiency of 47\% in a rectangular cavity.
\end{abstract}

\maketitle

%\newpage
%{\bf Keywords}: cavity, resonant energy transfer, upconversion, organic, exciton, photon, quantum electrodynamics

%INTRODUCTION
%%%%%%%%%%%%%%%%%%%%%
\section{Introduction}
The utility of a given source of light can be extended by transforming it to a higher frequency through processes collectively known as upconversion. This can be exploited to activate medicine at targeted locations within the body\citep{Ang_2011,Chatterjee_2010, Theranostics_Chen_2014, Cancer_Dai_2013}, increase the efficiency of solar energy harvesting\citep{Xie_2012,Zou_2012,Trupke_2006}, or even to increase the growth rate of plants for biofuels\cite{Wondraczek_2013}. It may well be that up/down conversion, within the strong coupling limit, can also be incorporated into emerging quantum information technologies. 

Energy pooling is a particular type of upconversion in which two donor molecules are separately excited by absorption followed by a simultaneous energy transfer to an acceptor molecule. It has been experimentally observed in a fluorescein-donor/stilbene-acceptor system\cite{Nickoleit_1997}, and a perturbative, quantum electrodynamics framework for the process was subsequently established\citep{Jenkins1998,Jenkins1999}. This, in turn, was used to computationally estimate the rate of pooling, alongside those of competing relaxation pathways, for the fluorescein/stilbene system as well as a new hexabenzocoronene(HBC)/oligothiophene assembly\cite{LaCount_2015}. It was demonstrated that energy pooling can be much greater than other competing processes, and a set of design rules were laid out for the requisite properties of ideal donor/acceptor pairs. Very recent experimental efforts adopted rhodamine-6G and stilbene-420, which have many of these properties, to successfully elicit a relatively high pooling efficiency~\cite{Weingarten_2017}.

A major obstacle to realizing efficient energy pooling is the requirement that two donor molecules be simultaneously excited while in close proximity to each other. This typically causes the rate of spontaneous donor emission to surpass the rate at which pairs are excited. The problem can be overcome by using light of sufficiently high intensity~\cite{Weingarten_2017} but demanding this level of illumination makes it impractical for many of the applications envisioned.

An intriguing and ultimately more practical approach is to extend the singlet lifetime of the donor molecules. This amounts to modifying the dominant pathway for de-excitation of the donor molecules. If that pathway is spontaneous emission (SE), then photonically confined environments can be designed to effectively increase exciton lifetimes. In general, such Inhibited Spontaneous Emission (ISE) occurs when the emission energy of a molecule falls within the photonic bandgap of a waveguide, cavity, or photonic crystal~\citep{Yablonovitch_1987, Flores_2002}. This has been exploited to more easily measure the magnetic moment of electrons\cite{Hanneke_2011}, create high quality single-photon sources for photonics\cite{Lund_2008}, and increase the exciton diffusion length in organic photovoltaic systems\cite{Kozyreff_2013}. It has also been used to extend the excited state lifetime of Rydberg atoms by a factor of twenty~\cite{Klepper_1981}. The consideration of ISE in association with energy pooling is new to the best of our knowledge.

\begin{figure}[t]\begin{center}
\includegraphics[width=0.48\textwidth]{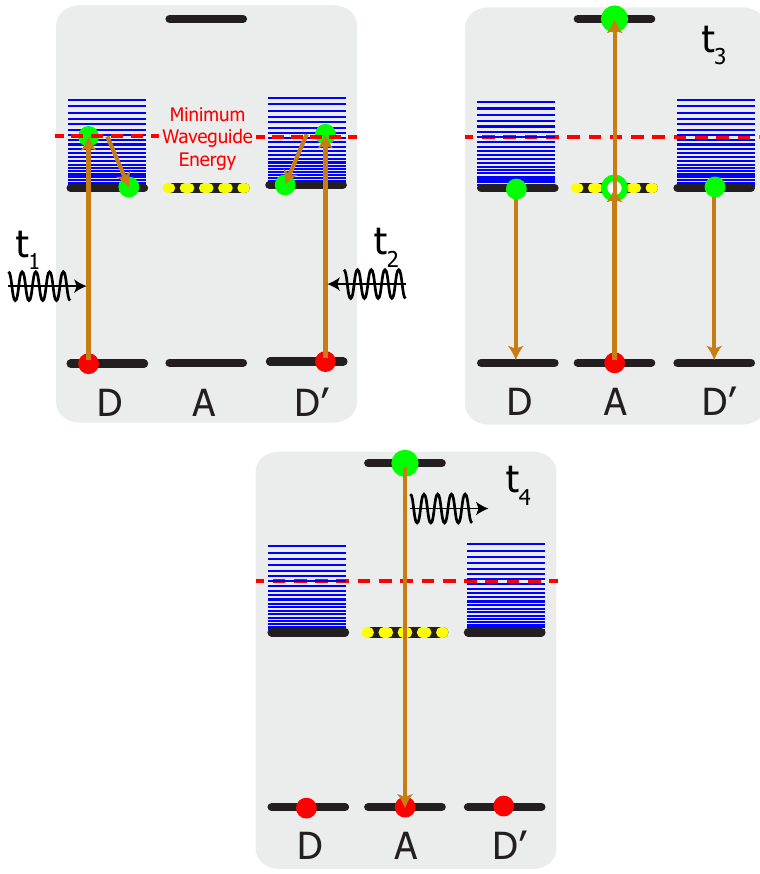}
\caption{\emph{Inhibited Spontaneous Emission via Stokes shift.} Energy level diagram for pooling molecules encapsulated within a geometry that exhibits a photonic bandgap. Top Left: Photons are absorbed by donors (D, D') at times $t_1$ and $t_2$ sufficiently close that excitons exist on both over a finite time interval. Top Right: Energy pooling results in the excitation of the acceptor (A), a simultaneous, three-body process at time $t_3$. Lower: Acceptor emission of a higher energy photon at time $t_4$.  Solid black lines indicate relevant excitonic energy levels, blue lines reflect the addition of phonons, and the yellow/black line is a virtual state.}
\label{Energy_Levels}
\end{center}
\end{figure}

SE can be inhibited in photonically constrained settings (PCS) as a result of naturally occurring Stokes shifts, as illustrated in Figure \ref{Energy_Levels}. The donor molecules absorb photons that are above the cutoff of the PCS, and the resulting exciton entangles with phonons~\cite{Zang_2015}. This causes a Stokes shift that reduces the exciton energy to below the waveguide cutoff, and SE is inhibited. As will be computationally demonstrated, even a modest Stokes shift is sufficient to substantially change the SE rate. 
 %
% FIGURE 1

The result is an exciton lifetime that is functionally similar to triplet-triplet annihilation (TTA), an upconversion method involving two triplet excitons.\cite{Baluschev_2006} This is significant because the primary advantage of TTA upconversion is the long lifetime of triplets which allows for efficient conversion even at low light intensity\cite{Ogawa_2015}. The increased longevity of singlet states afforded by ISE is not likely to exceed that which can be achieved via the forbidden transitions of triplet states~\cite{Hulet_1985}, but ISE-enhanced singlet pooling does offer an advantage over TTA upconversion. Triplets transfer energy through a Dexter process and so require small separations to achieve the requisite wave function overlap. ISE-enhanced singlets, though, can hop over longer distances via F{\"o}rster Resonant Energy Transfer (FRET). The use of ISE to improve the efficiency of such hops has been previously explored~\cite{Blum_2012}, where ISE was viewed as a means of modifying the local density of states. These  FRET rates are unaffected by the presence of the confining medium so long as the distance to the boundaries is larger than the separation between molecules\cite{LaCount_2016}.

The present work computationally explores the potential of ISE to increase energy pooling efficiency, the acceptor energy produced per unit of absorbed donor energy. A set of kinetic equations is developed that accounts for the primary competitors to energy pooling, most importantly  spontaneous emission from the donor and FRET from the acceptor back to the donor (back-FRET). Process rates are obtained from first principles analyses, and the entire kinetic model is embedded within an optimization routine. This allows molecule orientations and spacings to be tailored for the highest possible efficiency. The analysis procedure is applied to a well-studied donor/acceptor pair, and it is found that the pooling efficiency is substantially improved by exploiting ISE.

%
%Theory
%%%%%%%%%%%%%%%%%%%%%
\section{Theory}

The coupling between light and matter is assumed to be sufficiently weak that energy transfer processes can be considered within a perturbative Quantum Electrodynamics (QED) setting\cite{Andrews_2004}. The complete Hamiltonian can then be separated into independent light and matter contributions, $\hat{H}_0$, as well as a small light-matter coupling term, $\hat{H}_1$. Treating excitons as indivisible bosonic particles, the base Hamiltonian is then
\begin{eqnarray}
&&\hat{H}_0  = \hat{H}_{\mathrm {ex}} +  \hat{H}_{\mathrm {light}} , \quad
\hat{H}_{\mathrm {ex}}  = \sum_{j}\varepsilon_{j}\hat{c}^{\dagger}_{j}\hat{c}_{j}  , \nonumber \\
&&\hat{H}_{\mathrm {light}}=  \sum_{\lambda,\mathbf{k}} \hbar c k \,\hat{a}^{(\lambda, \mathbf{k})\dagger} \hat{a}^{(\lambda, \mathbf{k})} . 
\label{H0}
\end{eqnarray}
The purely excitonic component, $\hat{H}_{\mathrm {ex}}$, is in terms of the exciton annihilation operator, $\hat{c}_{j}$, of material state, $\ket{j}_{\mathrm {ex}}$, with bosonic commutation relations $[\hat{c}_{i}, \hat{c}^{\dagger}_{j}]_-=\delta_{ij}$. The excitonic energy  of molecule, $j$, is $\varepsilon_j$, and $c$ is the speed of light in vacuum. The photon component, $\hat{H}_{\mathrm {light}}$, is expressed in terms of the photon annihilation operator, $\hat{a}^{\mathbf{(\lambda, k)}}$, for which the modes are parametrized by vector $\bf k$ and polarizations $\lambda = 1, 2$. It destroys a photon in orientation and mode $(\lambda,\mathbf{k})$ and obeys the following bosonic commutation relations\cite{Andrews_SPIE_2013}:
\begin{equation}
[\hat{a}^{\mathbf{(\lambda, k)}}, \hat{a}^{(\gamma,\mathbf{p})}]_- = (8 \pi^3 \Omega)^{-1}\delta(\mathbf{k} -  \mathbf{p})\delta_{\lambda, \gamma} .
\end{equation}
Here $\Omega$ is a normalization volume. 

The interaction term is defined within the dipole approximation as $\hat{H}_1 = -\boldsymbol{\hat{\mu}}\cdot \mathbf{\hat{E}}$, where $\boldsymbol{\hat{\mu}} = e \hat{\bf{r}}$ is the electric dipole moment operator that acts upon the excitonic states, and $\mathbf{\hat{E}}$ is the electric field operator that acts on the optical states. 

ISE relies on the existence of a photonic bandgap, so the environment must be confined in at least two directions. In planar waveguides, for instance, the TM$_{0}$ mode allows for EM-waves of any energy while the lowest modes TM$_{01}$ or TE$_{01}$  of rectangular waveguides have a minimum energy related to its dimensions.

Three types of confined settings were considered to determine the waveguide geometry that delivers the greatest improvement in pooling efficiency. The first configuration is a square cavity tuned to resonate with the absorption of the donor. This will minimize the energy lost through thermalization while still causing a decrease in the photoluminescence (PL) rate of the donors due to ISE. It will be referred to as a \emph{Square-Donor} waveguide. The second waveguide also has a square cross-section but is tuned so that the second-lowest energy level of the cavity is resonant with the lowest acceptor emission, referred to as a \emph{Square-Acceptor} waveguide. It will maximize the Purcell effect for acceptor emission and, depending on the energy levels of the encapsulated material, may still allow ISE to be exhibited by the donor. The third waveguide has a rectangular cross-section with one side tuned to donor absorption energy and the second to acceptor emission energy, a \emph{Rectangular} guide that is intended to elicit both a Purcell enhancement of the acceptor emission and ISE of the donor. The free-space case is also considered for the sake of reference.

The excitonic operator depends solely on the electronic properties of the molecules in the assumed weak-coupling setting, while the electric field operator depends only on the waveguide geometry. For free-space and rectangular cavities, this operator has the following forms~\cite{LaCount_2016}:
\begin{equation}
\mathbf{\hat{E}}_{\rm free}(\mathbf{r}, t) = \imath\sum_{\lambda,\mathbf{k}} \sqrt{\frac{\hbar ck}{2 \Omega \varepsilon}}
\hat{\mathrm{\mathbf e}}^{(\lambda,\mathbf{k})} e^{i\mathbf{k}\cdot\mathbf{r}}
{\mathrm e}^{\imath c k t}\hat{a}^{(\lambda, \mathbf{k})}+ \rm{H.c.} 
\label{Efree}
\end{equation}
and
%\begin{widetext}
\begin{eqnarray}
\mathbf{\hat{E}}_{\rm cav}(\mathbf{r}, t) &=& \imath\sum_{\mathbf{k}} \sqrt{\frac{\hbar ck}{\Omega \varepsilon}\frac{2 k_\eta^2}{k^2}}e^{i k_x x}
{\mathrm e}^{\imath c k t}
\left(\left({\rm sin}(k_y y){\rm sin}(k_z z)\hat{x}+\frac{}{}\right.\right. \\ \nonumber
& &	\left.\frac{i k_x k_y}{k_\eta^2}{\rm cos}(k_y y){\rm sin}(k_z z)\hat{y} - \frac{i k_x k_z}{k_\eta^2}{\rm sin}(k_y y){\rm cos}(k_z z)\hat{z} \right)\hat{a}^{TM,(\mathbf{k})}+ \\ \nonumber
& &	\left.\left(\frac{-i k k_z}{k_\eta^2}{\rm cos}(k_y y){\rm sin}(k_z z)\hat{y}+ \frac{i k k_y}{k_\eta^2}{\rm sin}(k_y y){\rm cos}(k_z z)\hat{z}
\right)\hat{a}^{TE,(\mathbf{k})}\right)
+ \rm{H.c.}
\label{Ecavity}
\end{eqnarray}
%\end{widetext}

%
The operator depends on the position vector, $\mathbf{r}$, and time, $t$. The $k_x$ and $k_y$ components of the wave vector are combined, $k_\eta^2 = k_y^2 + k_z^2 = \left( n \pi /a \right)^2 + \left( m \pi /b \right)^2$, $\varepsilon$ is the permittivity of free space, and $\hat{\mathrm{\mathbf e}}^{(\lambda,\mathbf{k})}$ are the orthonormal polarization vectors such that $\hat{\bf k} \cdot \hat{\mathrm{\mathbf e}}^{(\lambda,\mathbf{k})} = 0$. 

%
% FIGURE 2
\begin{figure}[t]\begin{center}
\includegraphics[width=0.48\textwidth]{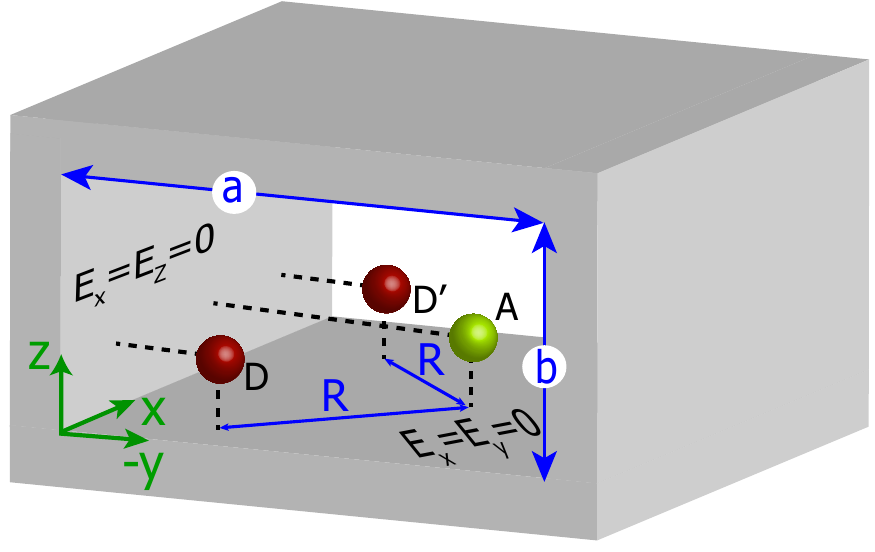}
\caption{Geometry and relative positions of molecules in a photonically confined setting.}
\label{Cavity_Geometry}
\end{center}
\end{figure}

As shown in the Appendix, this Hamiltonian can be subjected to perturbation theory to derive rate ($\Gamma$) expressions for absorption, photoluminescence (PL), internal conversion (IC), FRET, and energy pooling. Internal conversion was not calculated but was instead estimated using experimental values of fluorescent quantum yield combined with computed rates of SE.  Singlet-Singlet Annihilation (SSA) is an additional process of concern but can be thought of as a special case of FRET in which one donor transfers energy to the other, creating an excited state above the first excited one that undergoes a subsequent rapid decay into the first excited state. It is therefore implicitly accounted for in our approach.

\begin{figure}[t]\begin{center}
\includegraphics[width=0.48\textwidth]{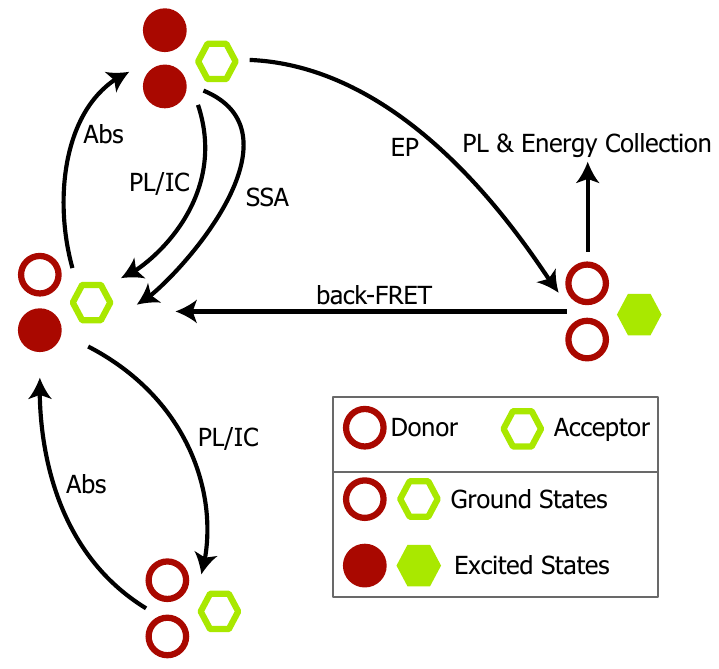}
\caption{\emph{Kinetic Model.} Schematic of energy transfer pathways considered: Absorption (Abs), Photoluminescence (PL), Internal Conversion (IC), Singlet-Singlet Annihilation (SSA), F{\"o}rster Resonant Energy from acceptor back to a donor (back-FRET), and Energy Pooling (EP).}
\label{Pathways}
\end{center}
\end{figure}

These rate relations can then be used to populate a kinetic model (Figure $\ref{Pathways}$), a set of coupled ordinary differential equations for concentrations, to quantify the efficiency, $\eta_{EP}$, of energy pooling relative to competing relaxation processes:

\begin{eqnarray}
& &	\Gamma_{Abs}^{(D_1)}[D_1^{S0}] + \Gamma_{FRET}^{(A\rightarrow D_1)}[D_1^{S0}][A^{S1}] + \Gamma_{FRET}^{(D_2\rightarrow D_1)}[D_1^{S0}][D_2^{S1}] = \\ \nonumber
& &	 (\Gamma_{PL}^{(D_1)}+\Gamma_{IC}^{(D_1)})[D_1^{S1}] + \Gamma_{FRET}^{(D_1\rightarrow D_2)}[D_1^{S1}][D_2^{S0}] + \Gamma_{SSA}^{(D_2\rightarrow D_1)}[D_1^{S1}][D_2^{S1}] + \Gamma_{EP}^{(D_1,D_2\rightarrow A)}[D_1^{S1}][D_2^{S1}][A^{S0}]
\label{kineticD1}
\end{eqnarray}
\begin{eqnarray}
& &	\Gamma_{Abs}^{(D_2)}[D_2^{S0}] + \Gamma_{FRET}^{(A\rightarrow D_2)}[D_2^{S0}][A^{S1}] + \Gamma_{FRET}^{(D_1\rightarrow D_2)}[D_1^{S1}][D_2^{S0}] = \\ \nonumber
& &	(\Gamma_{PL}^{(D_2)}+\Gamma_{IC}^{(D_2)})[D_2^{S1}] + \Gamma_{FRET}^{(D_2\rightarrow D_1)}[D_1^{S0}][D_2^{S1}] + \Gamma_{SSA}^{(D_1\rightarrow D_2)}[D_1^{S1}][D_2^{S1}] + \Gamma_{EP}^{(D_1,D_2\rightarrow A)}[D_1^{S1}][D_2^{S1}][A^{S0}]
\label{kineticD2}
\end{eqnarray}
\begin{equation}
\Gamma_{EP}^{(D_1,D_2\rightarrow A)}[D_1^{S1}][D_2^{S1}][A^{S0}] = (\Gamma_{PL}^{(A)}+\Gamma_{IC}^{(A)})[A^{S1}] + \Gamma_{FRET}^{(A\rightarrow D_1)}[D_1^{S0}][A^{S1}] + \Gamma_{FRET}^{(A\rightarrow D_2)}[D_2^{S0}][A^{S1}]
\label{kineticA}
\end{equation}
\begin{equation}
1 = [D_1^{S0}]+[D_1^{S1}] = [D_2^{S0}]+[D_2^{S1}] = [A^{S0}]+[A^{S1}]
\label{conc}
\end{equation}
\begin{equation}
\eta_{EP} = \frac{\Gamma_{PL}^{(A)}[A^{S1}]E_1^{A}}{\Gamma_{Abs}^{(D_1)}[D_1^{S0}]E_1^{D_1}+\Gamma_{Abs}^{(D_2)}[D_2^{S0}]E_1^{D_2}} .
\label{eff}
\end{equation}

Here $[\xi^\zeta]$ represents the concentration of molecule $\xi$ in state $\zeta$. It was assumed that the relative concentration of donors to acceptors was 2:1, as one might expect from a 3-body process. The use of concentrations to form the kinetic model sacrifices information about local proximity in favor of statistical averages. While simplifying the problem, such a coarse-grain model cannot capture effects associated with a non-uniform distribution of molecules. All concentrations are normalized such that there is only one of each molecule per unit volume (Eq. \ref{conc}). $E_1^{D_1}$ and $E_1^{D_2}$ are the excitation energies for donor one and donor two, and $E_1^{A}$ is the emission energy of the acceptor.

%
% FIGURE 3

This kinetic model can be used to quantify the steady-state efficiency, $\eta_{EP}$, here defined as the energy emitted by the system per energy unit absorbed (Eq. \ref{eff}). The system of equations is too complicated to obtain a general analytic result, so a numerical approach was taken (vide infra) for a specific set of donor and acceptor molecules: fluorescein isothiocyanate (FITC) donors and a hexabenzocoronene (HBC) acceptor. Though not considered together, each molecule has been previously analyzed within a free-space setting with computational predictions for ground state properties as well as absorption and emission profiles found to match experimental measurements~\cite{LaCount_2015}. While the free-space pooling efficiency is too low for this pair to be technologically interesting, it is well-suited to demonstrate the potential benefit of ISE.  This is because FITC has a fluorescence quantum yield of $0.93$~\cite{Sjoback_1995}, and donors with a high  quantum yield (or a low rate of IC) are promising because it implies that SE may be a dominant relaxation pathway. The HBC acceptor also has features that serve to enhance the effect of ISE---a dense excited state manifold and rigid structure. Such molecular stiffness minimizes the energy loss through geometry reorganization, while the dense excited state manifold creates more quantum pathways for energy pooling to occur, thus increasing its rate. As we are concerned with the effect of ISE, we chose to neglect the IC of the acceptor. If its rate is not significantly greater than the SE rate of the acceptor, the efficiency of energy pooling will scale linearly with the fluorescence quantum yield of the acceptor. 

Even prior to a physically accurate numerical implementation, it is possible to artificially populate the kinetic equations and obtain a rudimentary sense of how ISE might influence pooling efficiency. To this end, the following rates were utilized: donor absorption of 10 kHz, donor IC  of 10 MHz, acceptor emission of 100 MHz, FRET of 10 GHz, and energy pooling of 100 GHz. It was assumed that the acceptor both absorbs and emits at twice the energy at which the donor emits,  $E^{A} = 2\left(E^{D}-\Delta E^{{\rm Stokes}}\right)$. The reduction in radiative lifetime comes directly from a prescribed Stokes shift, and the larger Stokes shift also means that less energy is emitted by the acceptor. The increase in Stokes shift drops the donor emission energy further below the minimum cavity energy, thus producing ISE.

\begin{figure}[t]\begin{center}
\includegraphics[width=0.40\textwidth]{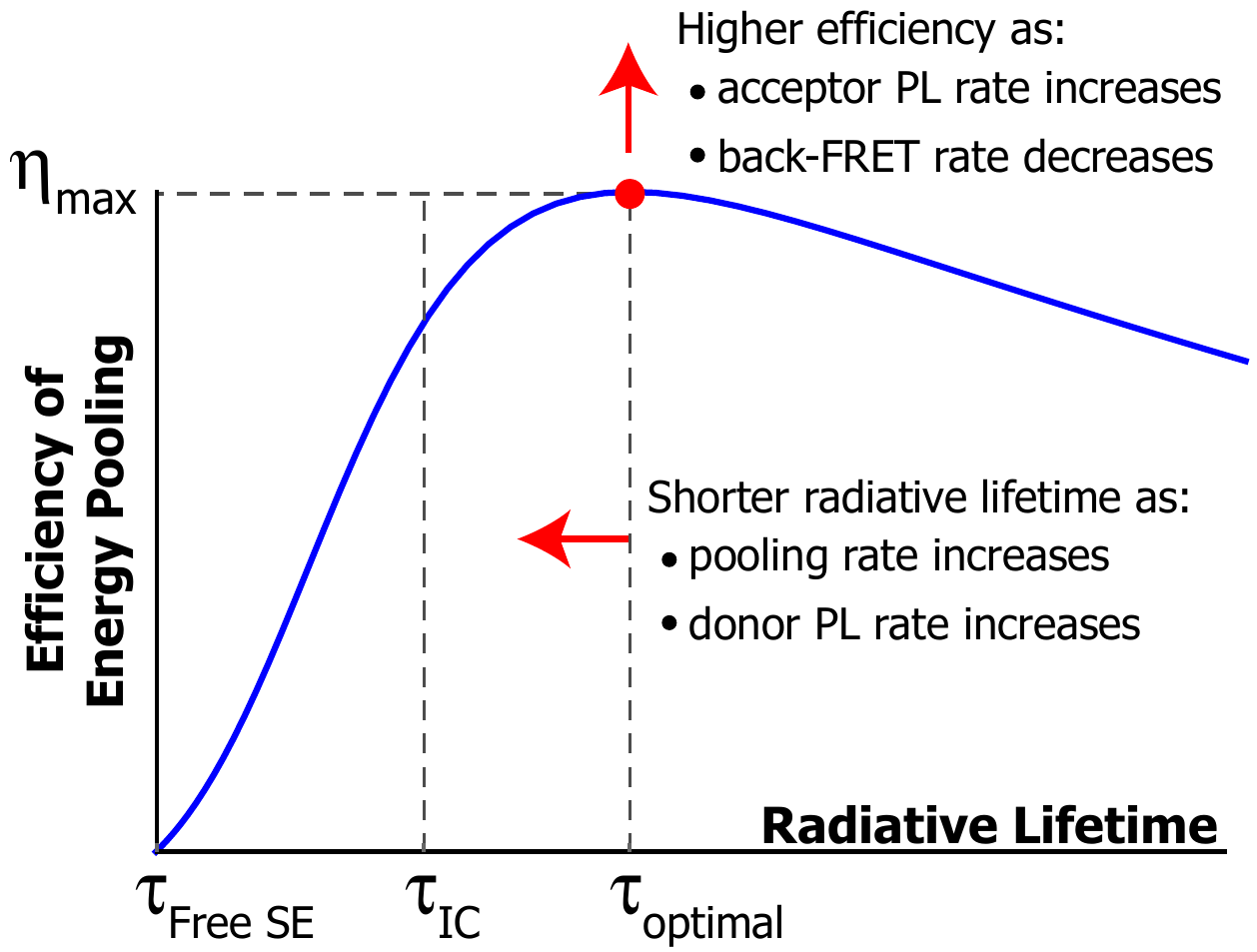}
\caption{\emph{Illustrative relation between radiative lifetime and pooling efficiency.} Linear-log plot demonstrating the efficiency gains of energy pooling as the radiative lifetime is extended using ISE. The parameters used are identified in the text.}
\label{Eff_Vs_SE}
\end{center}
\end{figure}

The resulting trends are captured in Figure $\ref{Eff_Vs_SE}$, where units are not shown because they are irrelevant in this qualitative analysis. The radiative lifetime was found by taking the inverse of the inhibited spontaneous emission rate which was varied by altering the Stokes shift of the donor. Therefore, increases in radiative lifetime were made indirectly by increasing the Stokes shift. As is graphically clear, there is a point at which the energy losses from the Stokes shift outweigh any increase in efficiency associated with an increase in donor lifetime. For the particular rates assumed, this happens for a Stokes shift of approximately one-third of the donor absorption energy.  Several factors were identified that influence both the peak efficiency value and its position. The peak efficiency can be increased by either increasing the SE rate of the acceptor or by decreasing the rate of back-FRET. These two processes compete, and the impact of the losses are quantified in the Results section that follows. The position of the peak can be shifted left (thereby reducing energy loss through the Stokes shift) by either increasing the rate of energy pooling or by increasing the rate of donor absorption. Increasing the rate of donor absorption causes an increase in excited donor concentration, and this shifts the peak to the left. Increasing the rate of energy pooling makes it more efficient, also resulting in a leftward shift of the peak. Additionally, a reduction in the donor emission line-width, $\gamma_{D}$, while holding the Stokes shift constant, causes the radiative lifetime to decrease. The line-width was taken to be the full-width half-maximum (FWHM) of the donor emission peak. Narrowing the emission peak will cause a smaller fraction of the emission to fall above the photonic bandgap of the cavity. 

Using the artificially populated kinetic model, an optimal Stokes shift was determined for a prescribed set of rates, excitation energy and donor emission line-width. These rates were allowed to increase or decrease within an order of magnitude, the excitation energy was allowed to range from 1 eV to 2 eV, and the donor emission line-width was varied over the range of 60-140 meV. It was observed that the optimal Stokes shift tended to fall within 20-30\% of the donor excitation energy. These qualitative relationships and rough efficiency estimates are useful in interpreting the numerical results of the specific calculations to follow.

%
% FIGURE 4

\section{Computational Details}
First principles calculations were used to obtain estimates for the parameters of the kinetics model comprised of Equations \ref{kineticD1} -- \ref{eff}. Geometry optimization and excited state analyses for HBC and FITC were carried out using the Q-Chem software package~\cite{QCHEM}. The HBC acceptor was modeled using a combination of Density Functional Theory (DFT) (ground state geometry) and Time-Domain Density Functional Theory (TD-DFT) (spectrum of excited singlet states). It is very rigid\cite{Kastler_2005}, so the Stokes shift was taken to be zero. A series of exchange-correlation functionals were considered, and the B3LYP\cite{B3LYP} functional was adopted because it predicted absorption properties closest to experimental spectra.  The TD-DFT analysis of HBC predicted a number of low-energy excited states that have no oscillator strength and do not appear in the absorption or emission spectra. These states were deemed unphysical and ignored.

A similar approach was attempted for FITC, but because it has a combination of local excitations and intramolecular charge transfer excitations, no exchange-correlation functional was found that accurately captures the properties of the molecule. It was therefore modeled using Spin-Opposite Scaling Second Order M\o{}ller-Plesset (SOS-MP2)\cite{SOSMP2} for the ground state geometry, configuration interaction singles and doubles (CIS(D))\citep{CISD1,CISD2} for excited state geometry, and Spin-Opposite Scaling Configuration Interaction Singles and Doubles (SOS-CIS(D))\cite{SOSCISD} for the spectrum of excited singlet states. While more computationally expensive than DFT/TD-DFT, this methodology was able to generate absorption and emission spectra consistent with experimental data.

The excited state calculations of FITC and HBC included the first 60 excited states of each molecule in both the ground state and excited state configurations. This was to ensure that any excited state within a range of $\sim$2.5 times the donor emission energy was included. Quantitatively, the cutoff ensured that the rate of energy pooling was converged to less than 1\% change per excited state included. The two-electron integral cutoff was set to $10^{-14}$ Ha, and the self-consistent field (SCF) convergence criteria for the electronic wavefunction was set to difference ratios between successive steps of $10^{-6}$ for HBC and $10^{-8}$ for FITC. The geometry optimization was considered converged when two of the following are satisfied: total of force magnitudes $< 0.000300$ a.u., total of displacement magnitudes $< 0.001200$ a.u., and total energy change $< 0.000001$ a.u.

\begin{figure}[t]\begin{center}
\includegraphics[width=0.48\textwidth]{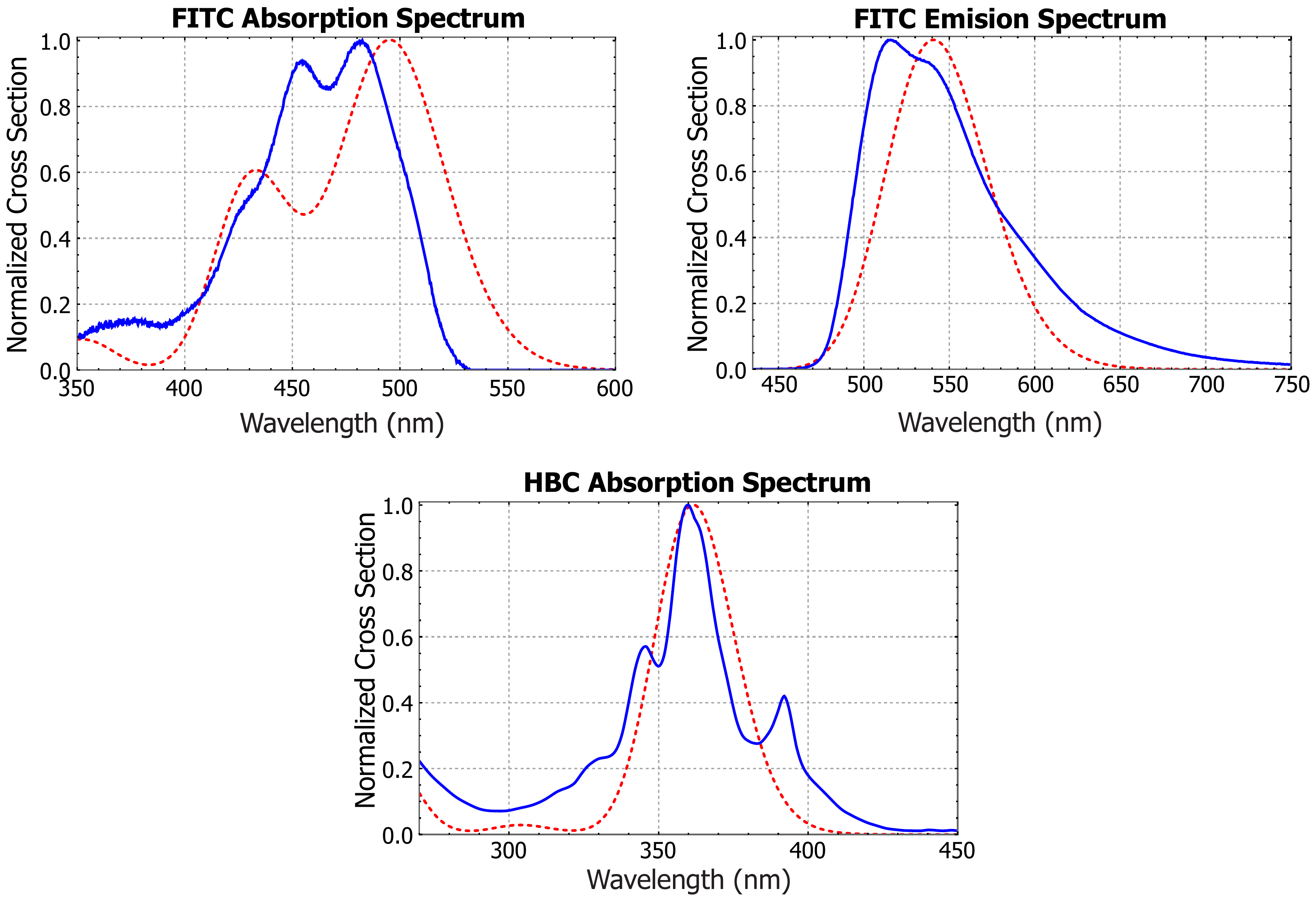}
\caption{\emph{Absorption Spectra.} Experimental\citep{Hu_2014,Du_1998,Dixon_2005,PhotochemCAD} (blue) and computed (red dashed) spectra of HBC and FITC.}
\label{Spectra}
\end{center}
\end{figure}

A comparison of the predicted and measured spectra for HBC and FITC is shown in Figure \ref{Spectra}. The computational data was subjected to a Gaussian convolution of the oscillator strength of each excited state\cite{Tiago_2006} for ease of comparison. A broadening of 121 meV was used for the absorption of FITC and 124 meV for the emission of FITC and the absorption/emission of HBC, empirically determined using the FWHM of the experimental data. These broadening parameters were also implemented in the kinetic model.
%
% FIGURE 5
\section{Results}
The set of kinetics equations, populated with DFT and TD-DFT data, can be numerically solved after specifying a specific orientation and spacing of the donor/acceptor triad.  This model was therefore embedded in an optimization routine which used maximum pooling efficiency for its objective function. The three molecules were treated as point dipoles and their relative orientation and separation distances were treated as free parameters. Each was assumed to be illuminated by the electric field found along the centerline of the waveguide (Figure \ref{Cavity_Geometry}). To be specific, the acceptor was positioned at (0,-a/2,b/2), the first donor at at (R,-a/2,b/2), and the second donor at (-R,-a/2,b/2).

%
% FIGURE 6
\begin{figure}[t]\begin{center}
\includegraphics[width=0.48\textwidth]{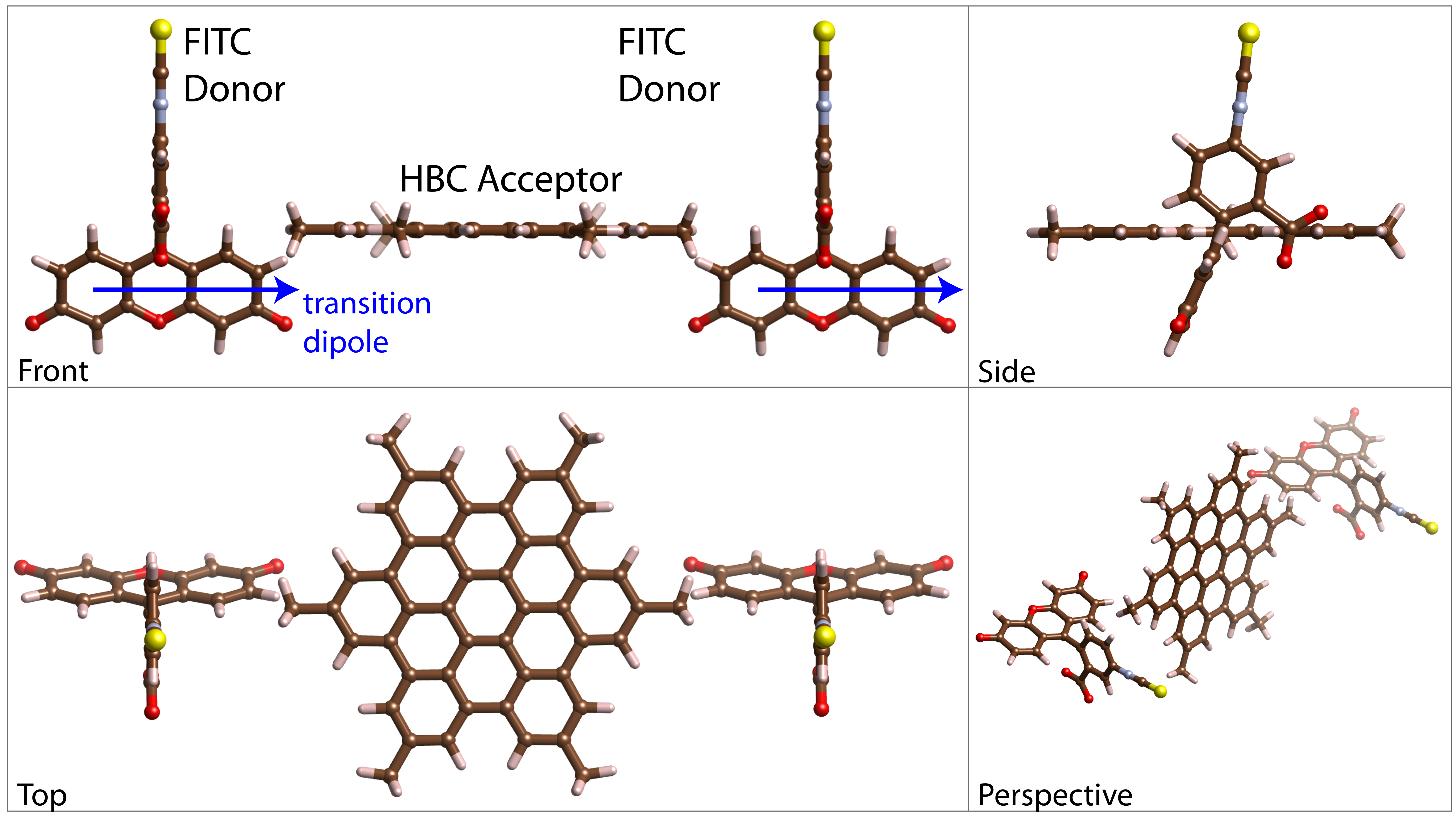}
\caption{\emph{Optimal Geometry and Orientation of FITC and HBC in Free Space.} Blue arrows indicate the direction of the lowest-energy transition dipole of the donors. The two-photon absorption tensor of the acceptor has nonzero elements only in the plane of the HBC.}
\label{molecules}
\end{center}
\end{figure}

Molecule orientations and positions were optimized in all three cavity settings, as well as free space, at an irradiance intensity of 1 MW/m$^2$. To give a physical correspondence, this is approximately 1000 suns, the highest irradiance used in concentrated photovoltaics. The optimal free-space geometry is shown in Figure \ref{molecules}. Optimal geometries for the photonically confined settings are visually indistinguishable with donor-acceptor separation distances 0.3 nm greater than that obtained for free space. The two-photon absorption (TPA) tensor for HBC has terms only in the plane of the molecule and is nearly isotropic, so that rotation of HBC within the plane had very little effect. In the optimal orientation, the FITC donors are diametrically opposed. Their emission transition dipoles are completely in the TPA-plane and are parallel to the axis separating the two donors. 

The rates and efficiencies of the processes in the free-space and cavity settings are summarized in Table \ref{FITC_HBC_rates}. The optimal configurations within each PCS result in a larger separation between donor and acceptors, lowering the rate of energy pooling. However, pooling competes with donor relaxation through SE and/or IC, so the negative impact of the decrease in pooling rate is minimal. Instead, the increase in separation lowers the rate of SSA along with back-FRET. The overall effect of this is an increase in pooling efficiency.

Among the three waveguide designs proposed, those with a square cross-section provided a nearly equal increase in efficiency. The Rectangular geometry gave the greatest increase in efficiency, benefiting from both Purcell enhancement of the acceptor emission and ISE of the donor.

% Table 1

\begin{table}[h]
\caption{Rates (Hz) and Efficiencies (\%) of FITC-HBC System}
\begin{centering}
\begin{tabular}{|c|c|c|c|c|c|c|c|c|}
\hline
Environment & $\Gamma_{Abs}^{(D)}$ & $\Gamma_{PL}^{(D)}$ & $\Gamma_{IC}^{(D)}$ & $\Gamma_{PL}^{(A)}$ & $\Gamma_{SSA}$ & $\Gamma_{FRET}^{(A\rightarrow D)}$ & $\Gamma_{EP}^{(D_1,D_2\rightarrow A)}$ & $\eta_{EP}$\tabularnewline
\hline 
\hline 
Free Space & 4.9\,E4 & 1.7\,E8 & 1.2\,E7 & 4.1\,E8 & 4.1\,E10 & 1.4\,E11 & 1.0\,E12 & 0.1\tabularnewline
\hline
Square-Donor & 1.4\,E4 & 7.6\,E6 & 1.2\,E7 & 1.3\,E9 & 1.0\,E10 & 3.7\,E10 & 6.0\,E10 & 0.8\tabularnewline
\hline 
Square-Acceptor & 7.3\,E3 & 9.5\,E6 &  1.2\,E7 & 1.8\,E9 & 1.4\,E10 & 4.8\,E10 & 1.1\,E10 & 0.8\tabularnewline
\hline 
Rectangular & 1.1\,E4 & 8.4\,E6 & 1.2\,E7 & 1.8\,E9 & 1.1\,E10 & 4.1\,E10 & 7.6\,E10 & 0.9\tabularnewline
\hline
\end{tabular}
\par\end{centering}
\label{FITC_HBC_rates}
\end{table}

\begin{figure}[t]\begin{center}
\includegraphics[width=0.40\textwidth]{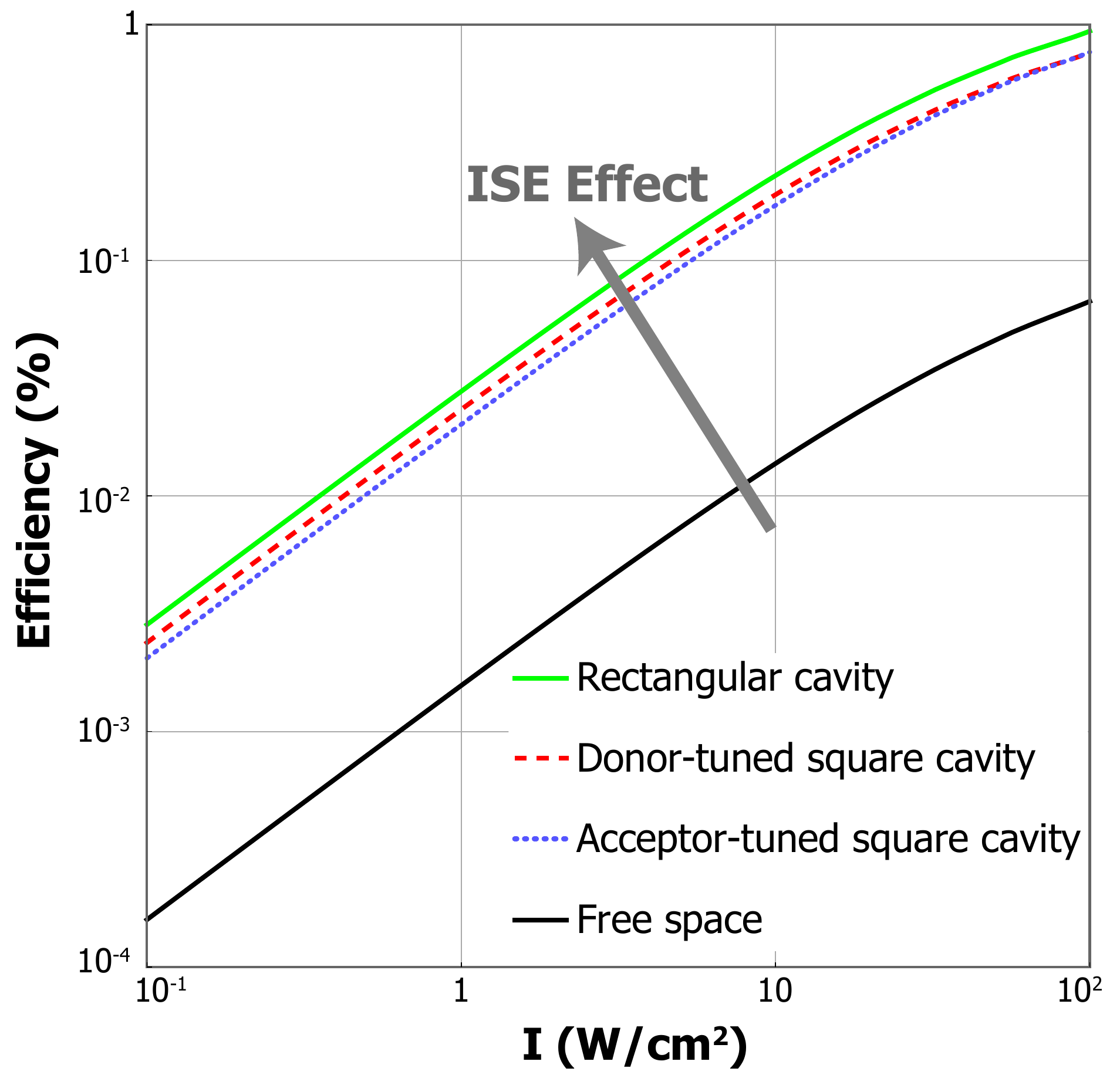}
\caption{\emph{Affect of ISE on Pooling Efficiency.} Energy pooling efficiency in free space (black), square-donor cavity (red/dark gray dashed), square-acceptor cavity (blue/light gray dashed), and rectangular cavity (green) for a range of 1-1000 suns.}
\label{intensity}
\end{center}
\end{figure}

With the optimal configuration fixed, the light intensity was subsequently varied as plotted in Figure \ref{intensity}. Not surprisingly, the pooling efficiency was found to increase at all light intensities considered. This is because, as the intensity increases, the excited donor concentration rises allowing  pooling to better compete against donor decay. However, the excited donor concentration eventually saturates, and further increases in intensity do not change the efficiency.

% FIGURE 7

These optimal geometries, and an excitation intensity of 1 MW/m$^2$, were then used to analyze the relative impact of competing relaxation pathways: donor IC, SSA, and back-FRET. The results are summarized in Table \ref{eff_study}. Donor IC was decreased such that the fluorescence quantum yield would rise from 0.93 to 0.99. Along the same lines, the SSA and back-FRET rates were decreased by a factor of 100. The resulting pooling efficiencies are only estimates because the geometry was kept fixed, but it is the trends that are important in any case. The largest pooling efficiency gains, in decreasing order of importance, are obtained by decreasing back-FRET, using a PCS to create ISE, reducing the rate of IC, and decreasing the rate of SSA. 

% Table 2

\begin{table}[h]
\caption{Pooling Efficiencies (\%) of FITC-HBC with Other Loss Mechanisms Dampened}
\begin{centering}
\begin{tabular}{|c|c|c|c|c|}
\hline
  & Free Space & Square-Donor & Square-Acceptor & Rectangular\tabularnewline
\hline 
\hline 
Baseline & 0.1 & 0.8 & 0.8 & 0.9\tabularnewline
\hline
Low SSA  & 0.1 & 0.9 & 0.9 & 1.1\tabularnewline
\hline 
Low IC & 0.1 & 1.2 & 1.2 & 1.5\tabularnewline
\hline 
Low back-FRET & 21 & 30& 29 & 30\tabularnewline
\hline 
Low back-FRET and SSA & 22 & 33 & 31 & 33\tabularnewline
\hline 
Low back-FRET and IC & 22 & 42 & 41 & 42\tabularnewline
\hline 
Low back-FRET, IC and SSA & 23 & 48 & 45 & 47\tabularnewline
\hline
\end{tabular}
\par\end{centering}
\label{eff_study}
\end{table}

Damping the SSA pathway offers only a minimal increase in pooling efficiency for all four environments. This is because the rate of SSA is significantly less than that for pooling in the optimized geometries. In other configurations, the losses through SSA may be more significant and its suppression therefore more important---e.g. arranging the three molecules so that they are equidistant in a plane, or if the spacing between donor molecules is less.

Increasing the fluorescence quantum yield causes the rate of donor IC to decrease, but the efficiency gains are once again very small in all settings. This is because IC is not relevant when radiative decay is fast and explains why the efficiency gains are highest in the PCSs.

The analysis shows that the damping of back-FRET has a substantial change on pooling efficiency,  increasing it  by  21\%, 29\%, 28\%, and 29\% for free space, Square-Donor, Square-Acceptor, and Rectangular environments, respectively. This is because, for the FITC/HBC system, back-FRET is much faster than SE from the acceptor. Even so, a PCS increases the pooling efficiency by approximately 9\%.

Blocking both SSA and back-FRET pathways increases the pooling efficiency to just over 30\% for both free-space and PCSs.  These numbers are impressive, but the highest theoretical pooling efficiency for the system is 69\%. It is the low excited donor concentration that causes the actual values to be much lower. This can be mitigated by either increasing the donor lifetime or by increasing the light intensity. 

Damping all three loss mechanisms would result in pooling efficiencies that are substantially affected by ISE. While the free-space efficiency is approximately 23\%, the PCSs have efficiencies in the range of 45\%---over 20\% higher. Stripping out all loss mechanisms but SE show how useful it can be to create a PCS for pooling.

\begin{figure}[t]\begin{center}
\includegraphics[width=0.40\textwidth]{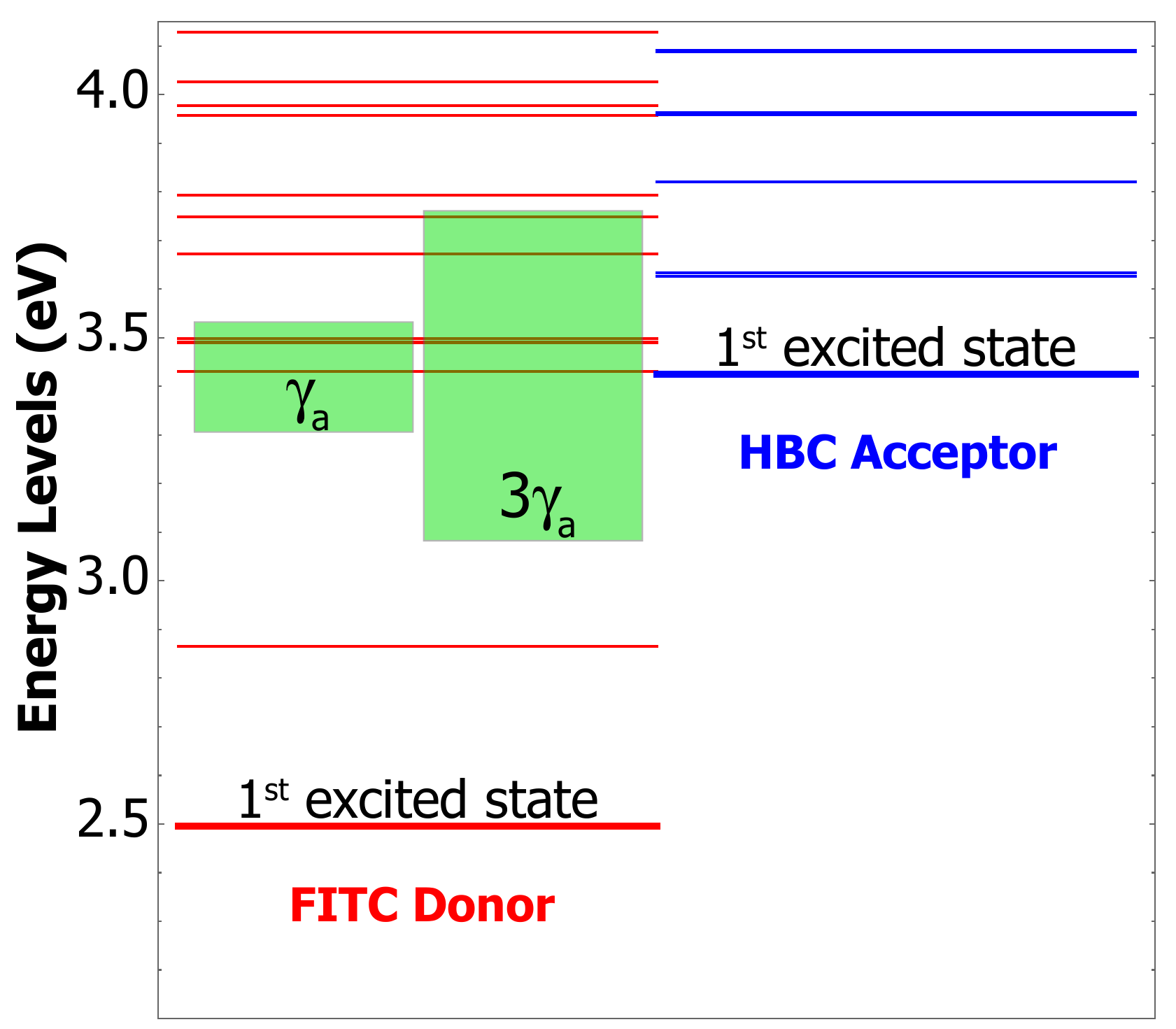}
\caption{\emph{Back-FRET Suppression.} Excited state energy levels of FITC donor (left, red) and HBC acceptor (right, blue). The shaded green bands highlight FITC states within $\gamma_{A}$ and $3\gamma_{A}$ of HBC emission from its first excited state. Removal of these donors levels suppresses back-FRET.}
\label{E_levels}
\end{center}
\end{figure}

For the FITC/HBC model system, the greatest loss mechanism is from back-FRET because it is significantly faster than acceptor PL as is clear from the following rate ratios: 300 (free space), 29 (Square-Donor), 27 (Square-Acceptor) and 23 (Rectangular). This can be mitigated by choosing a donor that does not have excited states in the energy range of the emission of the acceptor (Figure \ref{E_levels}). For instance, artificially removing excited donor states within a band of width equal to the acceptor broadening, $\gamma_{A}$ =120 meV, decreases the back-FRET-to-PL ratios to 13, 1, 1, and 1 while increasing the pooling efficiency to 2\%, 13\%, 13\% and 14\% for the free space, Square-Donor, Square-Acceptor, and Rectangular environments, respectively. Removing excited states within a band of width $3\gamma_{A}$ decreases the back-FRET-to-PL ratios to 3, 0.3, 0.3, and 0.2 and increases the pooling efficiency to 5\%, 22\%, 22\% and, 24\% for the free space, Square-Donor, Square-Acceptor, and Rectangular cavity environments, respectively.

%
% FIGURE 8

While the Stokes shift alone can be used to shift the emission to be within the photonic bandgap, it is also possible to tune the cavity so that the lowest energy mode of the cavity is resonant with a higher excited state on the donor. IC will rapidly lower excited donors to the first excited state. Such a large shift in energy puts the donor deeper within the photonic bandgap and causes ISE to be stronger. Of course, this comes at the expense of a larger loss of energy due to the initial IC, but the approach may still increase pooling efficiency. In the case of FITC, for instance, absorption to the second excited state (2.87 eV) actually gives a higher pooling efficiency, 2.0\% instead of 0.9\%. This is because the Stokes shift from the first excited state is only 0.21 eV, an 8.4\% shift in energy that is much lower than the 20---30\% ranged deemed to be optimal. This counterintuitive approach to optimizing pooling efficiency may be easily exploited for donors with the proper spacing between first and second excited states. 
%
%
%DISCUSSION
%%%%%%%%%%%%%%%%%%%%%
\section{Discussion}

Molecular assemblies composed of properly matched donor/acceptor pairs has been previously predicted to undergo measurable energy pooling upconversion~\cite{LaCount_2015}. That work resulted in a set of \emph{Molecular Design Criteria} intended to improve the rate of pooling without considering its overall efficiency within a complete kinetic model:
\begin{itemize}
\item{acceptor with large Two-Photon Absorption (TPA) cross section;}
\item{minimal spectral overlap of donor absorption and acceptor emission;}
\item{minimal spectral overlap of donor emission and excited donor absorption;}
\item{maximal spectral overlap of dual-donor emission and acceptor absorption, located as close as possible to the first excited state of the acceptor.}
\end{itemize}

A synergistic design strategy, explored in the current work, is to reduce the rate of competitive relaxation processes so as to increase the overall efficiency of pooling. In particular, SE was targeted and encapsulation within a photonically confined geometry was considered as a means of inhibiting this pathway. A donor Stokes shift subsequent to excitation reduces its energy to below the threshold for waveguide absorption. To quantify the impact of inhibiting SE, a kinetic model was developed and populated with first principles rate data based on perturbative quantum electrodynamics.

When populated with generic but typical rate values, the kinetic model showed that SE was indeed inhibited by increasing the donor Stokes shift of a hypothetical donor and that pooling efficiency will rise until the Stokes shift was about one-third the donor absorption energy. Beyond this level of relaxation, the energy losses from the Stokes shift overwhelm any advantage gained from reducing the rate of excited donor decay. Increasing the rate of acceptor emission or decreasing the rate of back-FRET were shown to increase the maximum theoretical efficiency. 

First principles data was then used to populate the kinetic model for a previously studied donor/acceptor pair---FITC donors and an HBC acceptor. Surprisingly, it was found that maximum efficiency is obtained by actually increasing the donor/acceptor separation over the optimal value for free space. This is because the losses associated with back-FRET reduce with increased separation, and this results in an overall efficiency gain even though the pooling rate is lower. The same would not be true for triplet upconversion which requires a high wavefunction overlap.

It was found that any of the three cavity designs considered resulted in a significant increase in energy pooling efficiency. However, a rectangular waveguide design produced the best results. One side was tuned for high donor absorption and ISE so that its length was L$_1=\pi \hbar c/E_1^{D}$. The second side was tuned for enhanced acceptor emission through the Purcell effect with a length of L$_2=\pi \hbar c/E_1^{A}$. Because the second side is always smaller than the first, it will also contribute to ISE.

The photonically confined geometries considered in this work were very simple. Settings such as nanopits or nanowires\cite{Bleuse_2011} are also possible so long as the dimensions are properly tuned and the confinement is in at least two Cartesian directions. Another possibility is to place the molecules within a photonic crystals. Nanogratings, which are only closed on three sides, do not exhibit the two-dimensional confinement required for ISE.

While ISE was found to improve pooling efficiency by roughly an order of magnitude, the efficiency was still less than 1\% for the FITC/HBC system studied. Subsequent analysis showed that this is because the rate of back-FRET is high due to substantial donor-absorption/acceptor-emission overlap. An absence of excited donor states within the effective emission range of the lowest excited acceptor state puts the rate of back-FRET on the same order as that of the desired acceptor emission. Within this setting, a free-space pooling efficiency of 23\% is shown to have an efficiency of 47\% in a rectangular cavity.

Utilizing molecules with minimal donor-emission/excited-donor-absorption overlap can further improve pooling efficiency by reducing the rate of SSA. However, this is of minor importance provided the donors are positioned on opposite sides of the acceptor as in the current study. The matching of donor-emission/acceptor-absorption energy levels is required for energy pooling to occur and is also important for minimizing the efficiency losses associated with acceptor relaxation to its lowest excited state~\cite{LaCount_2015}. The FITC/HBC system studied here does not meet this requirement as 25\% of the energy transferred to the acceptor is lost through acceptor internal conversion.

In summary, this computational study has shown that two additional Molecular Design Criteria should be added:
\begin{itemize}
\item{high emission quantum yield of donor;}
\item{a donor absorption/emission difference, on the order of 20-30\% the donor absorption energy, accomplished through a combination of Stokes shift and/or relaxation from a higher energy exciton state.}
\end{itemize}

The first criterion reflects the fact that there are both radiative and nonradiative relaxation pathways, but ISE only targets the former. It is therefore necessary to design a system where the latter is relatively small. The second ensures that donor relaxation reduces the emission energy sufficiently below the photonic cutoff to effectively inhibit SE. The shift range was estimated from a series of artificially populated kinetic models.

The combined designed criteria suggest that a pooling system composed of Rhodamine 101 donors and 1,4-Diphenylbutadiene acceptors is worthy of experimental consideration. Rhodamine 101 has a fluorescence quantum yield that is near unity and a weak absorption strength at twice the donor emission energy.\cite{Magde_1999}. The first absorption peak of 1,4-Diphenylbutadiene occurs at nearly twice Rhodamine 101 emission energy, it has a small Stokes shift and a higher quantum yield than similar molecules.\cite{Allen_1988}  A recent experimental study of similar but not ideal pairing, Rhodamine 6G (R6G) and Stilbene-420 (S420), reported efficiencies of 3.1-36\% in free space for light intensities of between 42000 and 720000 suns~\cite{Weingarten_2017}. Our design analysis suggests that encapsulating Rhodamine 101 and 1,4-Diphenylbutadiene in a photonically constrained setting will result in a significant enhancement in pooling efficiency at lower light intensities. 

On a device level, a relatively high concentration of acceptors to donors can be used to allow excited acceptors to transfer their energy to other acceptors in regions where there are no donors, so that energy loss through back-FRET is lessened. This is likely to be a significant factor in the high efficiency found in the R6G/S420 system with a blend ratio of 1:40~\cite{Weingarten_2017}. However, our concentration-based kinetic model cannot account for local density on donors relative to acceptors and so cannot accurately predict the effect of modifying the donor to acceptor ratio. We can posit a refinement, though, in which a solution of bonded donor-acceptor-donor (D-A-D) moieties are blended a solution of acceptors. When deposited as a film, the D-A-D molecules would perform energy pooling with upconverted energy efficiently transferred away to unbonded acceptors. This is another means of reducing back-FRET.

\section{Acknowledgements}
We are pleased to be able to acknowledge useful discussions with David Andrews on cavity quantum electrodynamics. All computations were carried out using the High Performance Computing facilities at the Colorado School of Mines.

\newpage

%APPENDIX: Derivation of Relaxation Rates

{\bf\appendix{APPENDIX: Derivation of Relaxation Rates}}

Several relaxation rate expressions are used in our kinetic model. In the free-space setting, these are fairly standard but are listed here for the sake of completeness. Their cavity counterparts, on the other hand, are subsequently derived.  

The rate of PL in free space is 
\begin{equation}
\Gamma_{PL, Free} = \frac{\left|\mathbf{\mu}^{(a,b)}\right|^2k^3}{3 \pi \hbar \varepsilon},
\label{PLfree}
\end{equation}
where $\mathbf{\mu}^{(b,a)}$ is the transition dipole moment from state a to state b, $k$ is the magnitude of the wave-vector of the photon being absorbed, $\varepsilon$ is the permittivity of the material.

The rate of absorption in free space is
\begin{equation}
\Gamma_{Abs, Free} = \sum_{n=1}^{\rm N}\frac{\pi I(k) \left|\mathbf{\mu}^{(n,0)}\right|^2}{3 \hbar c \varepsilon}\rho_f(E_n-\hbar c k),
\label{Absfree}
\end{equation}
where the sum over n is over all possible excited states, $I(k)$ is the irradiance of the light at the wave-vector k, $\rho_f$ is the density of final states, and $E_n$ is the energy of the n$^{\rm th}$ excited state.
The density of the excitonic states was taken to be a Gaussian distribution:
\begin{equation}
\rho_f(\Delta E) = \frac{1}{\sqrt{2\pi}\gamma}e^{-\frac{(\Delta E)^2}{2 \gamma^2}},
\label{gaussian}
\end{equation}
where $\gamma$ is a broadening term dominated by phononic effects. The value is empirically chosen for each molecular system based on the broadening observed in experimental absorption and emission spectra.

The cavity counterparts to these two rates are less standard. The PL rate associated with a rectangular cavity is found to be:
\begin{eqnarray}
\Gamma_{PL, Cavity} &=& \frac{2 c}{{\rm A} \varepsilon}\sum_{k_y,k_z}
\left(
	\left(\left|\mu_x^{(0,1)}\right|^2{\rm sin}(k_y y)^2 {\rm sin}(k_z z)^2k_\eta^2+\right.\right. \\ \nonumber
& &	\left.\left.(\left|\mu_y^{(0,1)}\right|^2 {\rm cos}(k_y y){\rm sin}(k_z z)k_z - \left|\mu_z^{(0,1)}\right|^2 {\rm sin}(k_y y) {\rm cos}(k_z z)k_y)^2\right)
	\int_{-\infty}^{\infty}\frac{\rho_f(\hbar c k-E_1)}{\sqrt{k_x^2+k_\eta^2}}dk_x+\right. \\ \nonumber
& &	\left.\left(\left|\mu_y^{(0,1)}\right|^2 {\rm cos}(k_y y)^2 {\rm sin}(k_z z)^2+\left|\mu_z^{(0,1)}\right|^2 {\rm sin}(k_y y)^2 {\rm cos}(k_z z)^2)\right)
	\int_{-\infty}^{\infty}\frac{k_x^2\rho_f(\hbar c k-E_1)}{\sqrt{k_x^2+k_\eta^2}}dk_x
\right),
\label{PLCavity}
\end{eqnarray}
where A is the cross sectional area of the cavity, and $k_\eta^2 = k_y^2+k_z^2$.

Following along the same lines, the rate of absorption in a rectangular cavity is:
\begin{eqnarray}
\Gamma_{Abs, Cavity} &=& \sum_{n=1}^N\frac{4 \pi I(k)}{\hbar c k^2 \varepsilon}\left(\left|\mu_x^{(n,0)}\right|^2{\rm sin}(k_y y)^2{\rm sin}(k_z z)^2k_\eta^2+
\left(\mu_y^{(n,0)}{\rm cos}(k_y y){\rm sin}(k_z z)\right.\right. \\ \nonumber
& &	\left.\left.	(k_x k_y-k k_z)+\mu_z^{(n,0)}{\rm sin}(k_y y){\rm cos}(k_z z)(k_x k_z-k k_y)\right)^2\right)\rho_f(E_n-\hbar c k)
\label{AbsCavity}
\end{eqnarray}

The rates of FRET and Energy Pooling have the same form in both free-space and cavity settings. The difference lies in the form of the dipole-dipole coupling tensor, $\mathbf{V}_{ij}$, but we have recently shown that it is reasonable to assume its free-space form holds in both settings~\cite{LaCount_2016}:
\begin{equation}
\mathbf{V}_{ij}(k,\mathbf{R}) = \frac{-e^{\imath k R}}{4\pi\varepsilon R^3}\left((1-\imath k R)(\delta_{ij}-3\hat{R}_i\hat{R}_j)+k^2R^2(\delta_{ij}-\hat{R}_i\hat{R}_j)\right) .
\label{Vij}
\end{equation}

The FRET expression is then\citep{Jenkins1999, Jenkins2004, Ford2014}
\begin{equation}
\Gamma_{FRET}^{(A\rightarrow B)} = \sum_{n=1}^N\frac{2 \pi}{\hbar}\left|\mathbf{\mu}_i^{B(n,0)}\mathbf{V}_{ij}(k^{(A)},\mathbf{R}_{AB})\mathbf{\mu}_j^{A(0,1)}\right|^2\rho_f(E_n^{B}-E_1^{A}),
\label{FRET}
\end{equation}
where $\mathbf{\mu}^{X(b,a)}$ is the transition dipole moment of molecule X from state a to state b, $E_n^{X}$ is the energy of the n$^{\rm th}$ excited state of molecule X, and $\mathbf{R}_{AB}$ is the displacement vector pointing from molecule A to molecule B.

Likewise, the energy pooling rate is\citep{Jenkins1999, Jenkins2004, Ford2014}
\begin{eqnarray}
\Gamma_{EP}^{(D_1,D_2\rightarrow A)} &=& \frac{2 \pi}{\hbar}\sum_{n=1}^N\left|
\mathbf{\mu}_i^{D_1(0,1)}\mathbf{V}_{ij}(k^{(D_1)},\mathbf{R}_{D_1A})\mathbf{\alpha}_{jk}^{A(n,0)}(k^{(D_1)},k^{(D_2)})\mathbf{V}_{kl}(k^{(D_2)},\mathbf{R}_{D_2A})\mathbf{\mu}_l^{D_2(0,1)}+\right. \\ \nonumber
& &	\mathbf{\mu}_i^{D_1(0,1)}\mathbf{V}_{ij}(k^{(D_1)},\mathbf{R}_{D_1D_2})\mathbf{\alpha}_{jk}^{D_2(0,1)}(k^{(D_1)},-k^{(D_1)}-k^{(D_2)})\mathbf{V}_{kl}(k^{(D_1)}+k^{(D_2)},\mathbf{R}_{D_2A})\mathbf{\mu}_l^{A(n,0)}+ \\ \nonumber
& &	\left.\mathbf{\mu}_i^{D_2(0,1)}\mathbf{V}_{ij}(k^{(D_2)},\mathbf{R}_{D_1D_2})\mathbf{\alpha}_{jk}^{D_1(0,1)}(k^{(D_2)},-k^{(D_1)}-k^{(D_2)})\mathbf{V}_{kl}(k^{(D_1)}+k^{(D_2)},\mathbf{R}_{D_1A})\mathbf{\mu}_l^{A(n,0)}\right|^2 \\ \nonumber
& &	\rho_f(E_n^{A}-E_1^{D_1}-E_1^{D_2})
\label{EP}
\end{eqnarray}
where $\mathbf{\alpha}_{jk}^{X(b,a)}$ is the two-photon transition tensor of molecule X from state a to state b:
\begin{equation}
\mathbf{\alpha}_{jk}^{\xi(fi)}(k_1,k_2) = \sum_\zeta \left(\frac{\mu_k^{\xi(f\zeta)}\mu_j^{\xi(\zeta i)}}{E_\zeta^\xi-E_i^\xi-\hbar c k_1+\imath\gamma}+\frac{\mu_j^{\xi(f\zeta)}\mu_k^{\xi(\zeta i)}}{E_\zeta^\xi-E_i^\xi-\hbar c k_2+\imath\gamma}\right)
\label{alphajk}
\end{equation}

\vspace{10 mm}

The rate of absorption in a rectangular waveguide can be derived using Fermi's Golden Rule:

\begin{equation}
\Gamma_{Abs, Cav} = \frac{2 \pi}{\hbar}\left| \bra{f}\hat{H}_1\ket{i} \right|^2\rho_f(E_f - E_i) = \frac{2 \pi}{\hbar}\left| \bra{n;0}\mathbf{\hat{\mu}}\cdot\mathbf{\hat{E}}\ket{0;\mathbf{k}}\right|^2\rho_f(E_n-\hbar c k),
\end{equation}

where for $\ket{A;{\bf B}}$ A represents the electronic state, and B represents the photonic state. Applying the electric dipole operator $\mathbf{\hat{\mu}}$ and the electric field operator $\mathbf{\hat{E}}$ to this expression gives:
\begin{eqnarray}
\label{Abs1}
\Gamma_{Abs, Cav} &=& \frac{2 \pi}{\hbar}\left| \sqrt{\frac{\hbar c k}{\Omega \varepsilon}\frac{2 (k_y^2+k_z^2)}{k^2}N_\gamma}
\left(\mathbf{\mu}_x {\rm sin}(k_yy){\rm sin}(k_zz)+\right.\right. \\ \nonumber
& & \left.\left.\mathbf{\mu}_y\frac{\imath(k_xk_y-k k_z)}{k_y^2+k_z^2}{\rm cos}(k_yy){\rm sin}(k_zz)+\mathbf{\mu}_z\frac{\imath(k_xk_z-k k_y)}{k_y^2+k_z^2}{\rm sin}(k_yy){\rm cos}(k_zz)
\right)
 \right|^2\rho_f(E_n-\hbar c k),
\end{eqnarray}
where $N_\gamma$ is the photon mode occupation number. An expansion of the terms results in the following rate for the rate of absorption within a rectangular cavity:
\begin{eqnarray}
\Gamma_{Abs, Cav} &=& \frac{\hbar c^2 k N_\gamma}{\Omega}\frac{4 \pi (k_y^2+k_z^2)}{\hbar^2c^2k^2\varepsilon}\left| 
\left(\mathbf{\mu}_x{\rm sin}(k_yy){\rm sin}(k_zz)+\frac{}{}\right.\right. \\ \nonumber
& & \left.\left.\mathbf{\mu}_y\frac{\imath(k_xk_y-k k_z)}{k_y^2+k_z^2}{\rm cos}(k_yy){\rm sin}(k_zz)+\mathbf{\mu}_z\frac{\imath(k_xk_z-k k_y)}{k_y^2+k_z^2}{\rm sin}(k_yy){\rm cos}(k_zz)
\right)
 \right|^2\rho_f(E_n-\hbar c k)
\end{eqnarray}
The first term on the right hand side is equivalent to the light intensity for a particular wave-vector.
\begin{eqnarray}
\Gamma_{Abs, Cav} &=& I(\mathbf{k})\frac{4 \pi (k_y^2+k_z^2)}{\hbar^2c^2k^2\varepsilon}\left| 
\left(\mathbf{\mu}_x{\rm sin}(k_yy){\rm sin}(k_zz)+\frac{}{}\right.\right. \\ \nonumber
& & \left.\left.\mathbf{\mu}_y\frac{\imath(k_xk_y-k k_z)}{k_y^2+k_z^2}{\rm cos}(k_yy){\rm sin}(k_zz)+\mathbf{\mu}_z\frac{\imath(k_xk_z-k k_y)}{k_y^2+k_z^2}{\rm sin}(k_yy){\rm cos}(k_zz)
\right)
 \right|^2\rho_f(E_n-\hbar c k).
\end{eqnarray}

A similar approach is used to derive the rate of spontaneous emission in a rectangular waveguide. The initial form is similar in form to Eq. \ref{Abs1}:
\begin{eqnarray}
\Gamma_{PL, Cav} &=& \frac{2 \pi}{\hbar}\sum_\mathbf{k}\left| \sqrt{\frac{\hbar c k}{\Omega \varepsilon}\frac{2 (k_y^2+k_z^2)}{k^2}}
\left(\mathbf{\mu}_x{\rm sin}(k_yy){\rm sin}(k_zz)+\right.\right. \\ \nonumber
& & \left.\left.\mathbf{\mu}_y\frac{\imath(k_xk_y-k k_z)}{k_y^2+k_z^2}{\rm cos}(k_yy){\rm sin}(k_zz)+\mathbf{\mu}_z\frac{\imath(k_xk_z-k k_y)}{k_y^2+k_z^2}{\rm sin}(k_yy){\rm cos}(k_zz)
\right)
 \right|^2\rho_f(E_n-\hbar c k)
\end{eqnarray}
Expanding this expression, and write the $k_x$ component of the sum as an integral gives
\begin{eqnarray}
\Gamma_{PL, Cav} &=& \frac{4 \pi c}{\Omega \varepsilon}\sum_{k_y,k_z}\frac{L_x}{2\pi}\int_0^\infty dk_x\frac{k_y^2+k_z^2}{k}\left|
\left(\mathbf{\mu}_x{\rm sin}(k_yy){\rm sin}(k_zz)+\right.\right. \\ \nonumber
& & \left.\left.\mathbf{\mu}_y\frac{\imath(k_xk_y-k k_z)}{k_y^2+k_z^2}{\rm cos}(k_yy){\rm sin}(k_zz)+\mathbf{\mu}_z\frac{\imath(k_xk_z-k k_y)}{k_y^2+k_z^2}{\rm sin}(k_yy){\rm cos}(k_zz)
\right)
 \right|^2\rho_f(E_n-\hbar c k) .
\end{eqnarray}
The normalization volume $\Omega$ and the normalization length in the x-direction $L_x$ are simplified to the cross-sectional area of the waveguide $A$. The result is that the rate of PL in the rectangular cavity is
\begin{eqnarray}
\Gamma_{PL, Cav} &=& \frac{2 c}{A \varepsilon}\sum_{k_y,k_z}\int_0^\infty dk_x\frac{k_y^2+k_z^2}{k}\left|
\left(\mathbf{\mu}_x{\rm sin(k_yy)}{\rm sin}(k_zz)+\right.\right. \\ \nonumber
& & \left.\left.\mathbf{\mu}_y\frac{\imath(k_xk_y-k k_z)}{k_y^2+k_z^2}{\rm cos}(k_yy){\rm sin}(k_zz)+\mathbf{\mu}_z\frac{\imath(k_xk_z-k k_y)}{k_y^2+k_z^2}{\rm sin}(k_yy){\rm cos}(k_zz)
\right)
 \right|^2\rho_f(E_n-\hbar c k)
\end{eqnarray}
From here the density of states Eq. \ref{gaussian} is substituted and the integral is computed numerically.

\newpage
%\bibliography{ISE}
\providecommand{\latin}[1]{#1}
\providecommand*\mcitethebibliography{\thebibliography}
\csname @ifundefined\endcsname{endmcitethebibliography}
  {\let\endmcitethebibliography\endthebibliography}{}

\newpage

\begin{figure}[h]
\centering{}
\includegraphics[width=0.50\textwidth]{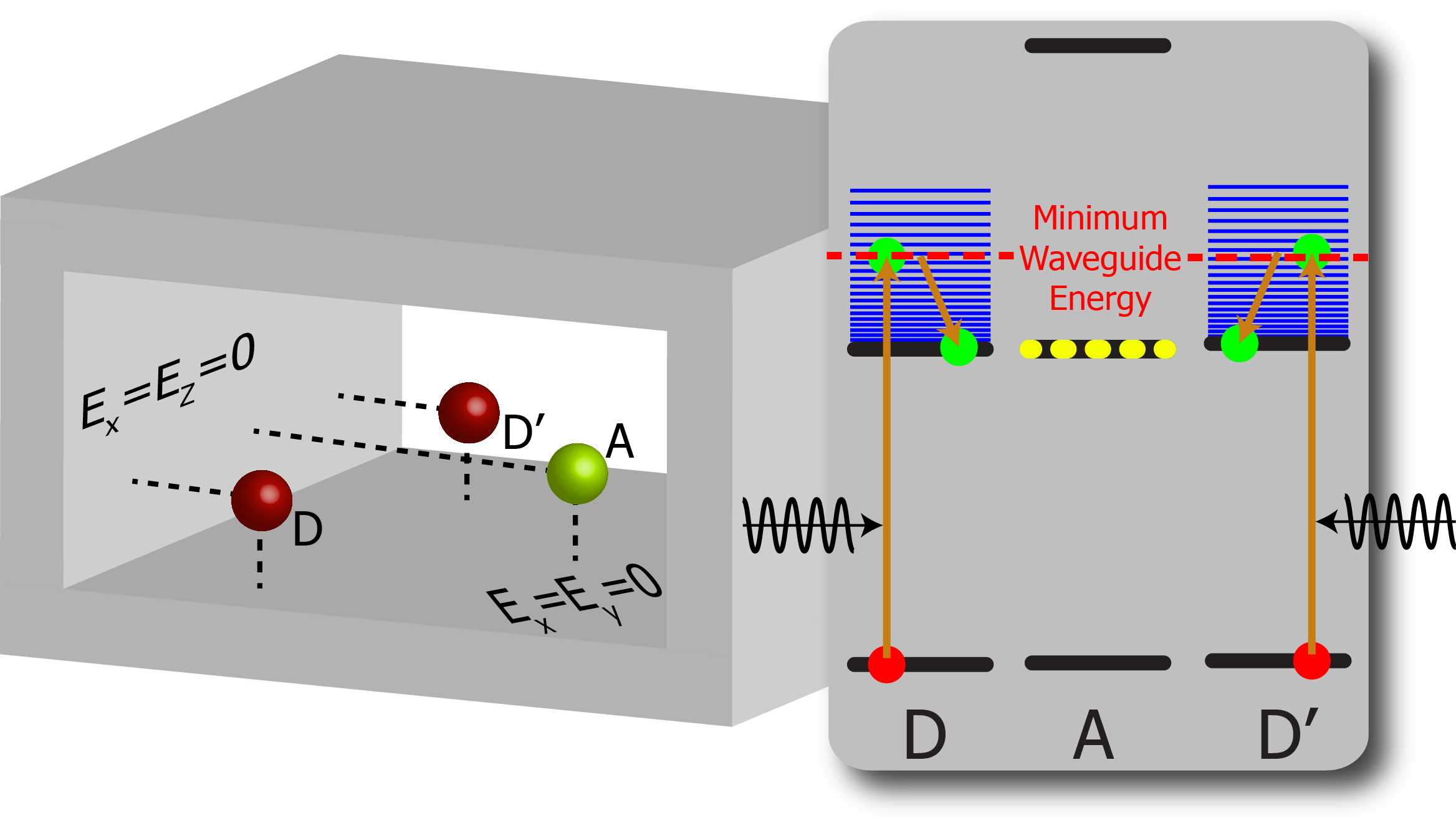}
\end{figure}

\newpage

\end{document}